 \documentstyle[12pt]{article}

\let\a=\alpha \let\b=\beta  \let\d=\delta \let\e=\epsilon
  \let\q=\theta  
 \let\m=\mu    
\let\s=\sigma   \let\f=\phi \let\c=\chi \let\y=\psi
      \let\G=\Gamma   
    \let\F=\Phi \let\Y=\Psi
 
\let\la=\label \let\ci=\cite 
  
 \def\bd{\begin{document}} \def\ed{\end{document}}
\def\ds{\documentstyle} \let\fr=\frac \let\bl=\bigl \let\br=\bigr
\let\Br=\Bigr \let\Bl=\Bigl
\let\bm=\bibitem
\let\na=\nabla
\let\pa=\partial \let\ov=\overline
\newcommand{\be}{\begin{equation}}
\newcommand{\ee}{\end{equation}}
\def\ba{\begin{array}}
\def\ea{\end{array}}
\newcommand{\ho}[1]{$\, ^{#1}$}
\newcommand{\hoch}[1]{$\, ^{#1}$}
\newcommand{\bea}{\begin{eqnarray}}
\newcommand{\eea}{\end{eqnarray}}
\newcommand{\ra}{\rightarrow}
\newcommand{\lra}{\longrightarrow}
\newcommand{\Lra}{\Leftrightarrow}
\newcommand{\ap}{\alpha^\prime}
\newcommand{\bp}{\beta^\prime}
\newcommand{\tr}{{\rm tr} }
\newcommand{\Tr}{{\rm Tr} }
\newcommand{\NP}{Nucl. Phys. }
\newcommand{\tamphys}{\it Center for Theoretical Physics\\
Physics Department \\ Texas A \& M University
\\ College Station, Texas 77843}

\begin{document}

\hfill{CTP-TAMU-69/92; HEP-TH 9210092}

\vspace{24pt}

\begin{center}
{ \large {\bf A Natural Mechanism for Supersymmetry Breaking
with Zero Cosmological Constant\footnote{Revised and Corrected Version} }}

\vspace{36pt}

J. A. Dixon

\vspace{6pt}

{\tamphys}

\vspace{6pt}

\vspace{6pt}

\underline{ABSTRACT}

\end{center}

Supersymmetric theories in four dimensions with chiral superfields have very
rich BRS cohomology, which gives rise to potential anomalies in theories that
contain composite antichiral spinor superfields.  Assuming the coefficients are
non-zero, absence of the anomalies would generate new constraints on theories.
In addition,  the anomalies give rise to a new kind of supersymmetry breaking
which is quite different from the known kinds, and also naturally yields a zero
cosmological constant after supersymmetry breaking.

\vfill

\baselineskip=24pt

\pagebreak

\setcounter{page}{1}

\section{Introduction}

There are some major problems in superstring theory that call for
some new ideas.  What is special about four dimensions of spacetime?
What picks out the standard model at low energy?
Why is supersymmetry broken?  Why is the cosmological constant zero
after supersymmetry breaking?

Some recent results in the BRS cohomology of N=1, D=4 supersymmetry
suggest that the answers to some of these questions may be hidden in
the subject of supersymmetry anomalies.  The purpose
of this article is to discuss the possibilities, and to try to point
out what is known, what seems possible, and what is unknown but worth
finding out.

One of the remarkable features of N=1, D=4
global supersymmetry is that when it is
unbroken it naturally gives a zero energy density to the vacuum.
   Another feature that looks
promising is that this is not spoiled by the spontaneous breaking
of gauge and internal symmetries, which happens naturally and
in many interesting ways in globally supersymmetric theories.

However, when global supersymmetry is spontaneously
broken by the Fayet-Iliopoulos \ci{Fayet} or O'Raifeartaigh
\ci{O'R} mechanisms,
the vacuum must acquire an energy density.

The unfortunate consequence of this is
that  while unbroken supersymmetry is not observed,
so that supersymmetry must somehow be broken, the
vacuum energy density generated when supersymmetry
is spontaneously broken generates
a cosmological constant that is far too large to be
consistent with the experimental size of the universe.
And when global supersymmetry is extended to
supergravity and superstring theories,
this problem still tends to be present in the sense that a zero
cosmological term requires an `unnatural' fine tuning of the parameters
of the theory.

A second problem of spontaneous supersymmetry breaking is that it is rather
hard to achieve.  Models tend
to have a contrived look, with invariant chiral superfields
or Abelian gauge groups needed. For phenomenological
purposes, non-Abelian theories with simple gauge groups are more appealing
in many ways.  Furthermore, in global supersymmetry,
 there are sum rules regarding masses
of broken supermultiplets that give trouble with phenomenology.
Is there a way to break supersymmetry that is not `spontaneous'
or explicit?

As is well known, freedom from gauge and gravitational
anomalies in the heterotic string puts stringent requirements
on the possible Yang-Mills gauge group in ten space-time dimensions.
Are there any more anomalies than these?

The existence of anomalies is governed by the BRS cohomology of the
relevant symmetry.  This is entirely unknown for any supersymmetric theory
except rigid N=1, D=4 supersymmetry, where partial results have been
obtained \ci{prl}  \ci{cmp2}
after a decade during which incorrect results \ci{pss}
were generally
accepted.  The incorrect results stated that the cohomology was trivial--
no new anomalies were possible.  The correct results indicate that the
cohomology is  very complex in D=4.  It is entirely possible that for D=10
supersymmetry, there are anomalies that render the theory inconsistent.
This kind of result could account for the observed dimension of space-time.
The calculation of the BRS cohomology for this case is a very non-trivial
task however.  It is also possible that the D=4 superstring may be
subject to higher order constraints  relating to the presently known
cohomology of supersymmetry.  To determine this one needs to extend the
known results to local supersymmetry.   This requires a careful demonstration,
probably in component form,
that antichiral spinor superfields couple to Poincare supergravity.
It is shown in \ci{west} that they do not couple to anti-de-Sitter
supergravity.
When this is done, it is probably straightforward to show that the rigid
cohomology carries over to the local case.
Let us assume that both of these work out.

In this situation,
one can  construct an explanation of supersymmetry breaking which
accounts for the zero cosmological constant in a natural way.
Let us assume that we have a four dimensional locally
supersymmetric effective action (derived from the superstring)
 which is free of anomalies,
including supersymmetry anomalies--this may be a very stringent
requirement of course.

Then we introduce composite gauge-invariant antichiral spinor
superfield operators to the theory
by coupling them to external superfield sources.
The BRS cohomology states that these new terms may be subject to
anomalies.  Recall that the observable states in our theories are
in fact usually composite and gauge-invariant.
For example the proton is constructed from
quarks, and the electron and photon are also composite in the sense
that one must write them as composite using a Higgs field.  The
VEV of the Higgs allows a `single particle' to appear from the composite
field.  These operators do not normally appear in the action.

It appears from the BRS cohomology of supersymmetry in D=4 that
all these composite operators are subject to supersymmetry anomalies.
So we would not expect to observe supersymmetry in the
masses and interactions of the corresponding `bound states'.
On the other hand, because the fundamental theory
is anomaly free, the theory is consistent and unitary.
So this is  a new mechanism of supersymmetry breaking, and we could
envisage using it alone without any spontaneous supersymmetry breaking
at all.

The appealing feature of this is that this mechanism of supersymmetry breaking
leaves the vacuum energy untouched.  Now the vacuum energy
is normally zero
in supertheories that come out of the superstring. It is still zero
after spontaneous breaking of gauge symmetries,
which naturally occurs in supersymmetric theories as noted above,
Hence, using the superanomaly mechanism above, the
{\em cosmological constant is naturally zero after gauge and supersymmetry
breaking.}

It is clear from the BRS cohomology that a very complicated
supersymmetry breaking would indeed be induced in this way.
Moveover {\em supersymmetry breaks itself}--the breaking occurs in all
supersymmetric theories and it is
not a consequence of special tinkering with the theory to break it
so that it can match experiment.

So where is the `catch'?  The catch is that
it is not yet known how to compute the coefficients
of these anomalies--admittedly, if these are all zero, that is a
very big catch indeed.  It is probable that non-zero
coefficients do not appear in one-loop calculations.  However,
despite the rough arguments in \ci{prl}
to the effect that all the coefficients should be zero,
I now believe that they do appear in higher orders of perturbation theory.

This is essentially because there does not appear to exist
a gauge-invariant and supersymmetric regularization procedure
for supersymmetric theories.
It would be hard to define one without
simultaneously eliminating the known axial anomalies, which
would contradict the accepted view that these are present.
Evidently this is a delicate problem.

Obviously it is somewhat unsatisfactory to offer physical
interpretations before finding the coefficients.
The excuse offered is that one needs a motivation to
do the tricky calculations of the anomalies, and also of the
BRS cohomology of rigid and local supersymmetry in various dimensions,
and this paper is designed to supply that motivation.

Several objections might be made not counting the above catch.
\begin{enumerate}
\item
The first objection is that it would be strange to have observable
states that are not supersymmetric when the underlying theory is
supersymmetric.  I agree but do not yet see that it is also impossible--
in fact it is rather similar in some ways to
spontaneous symmetry breaking.
\item
It is totally unclear whether the splitting obtained in this way
would match experiment, and that seems to involve an interesting and
complex mixing problem.   But presumably the predictions are
at least testable.
\item
Is it reasonable to state that the observable states of a theory
are not the fundamental fields of the theory, but are actually
composite operators?  I think this is natural in the context of
the standard model and our accepted beliefs about QCD and spontaneous
breaking of gauge symmetry.
\item
The supersymmetry anomalies probably occur only beyond one loop and
they probably have higher order corrections.  They also seem to involve
mass parameters in general. This is all quite different from
what happens for the known chiral anomalies.  The chiral anomalies
do not get renormalized because they are connected to the index theorem.
Hence we might not expect an index theorem interpretation
for the present anomalies.
\item
This might also raise
worries that the effects are too small to account for phenomenology.
However, since the anomalies probably involve mass parameters as well
as coupling constants, there is hope that large mass splittings can be
generated.
\end{enumerate}

Now let us get to facts and leave speculation.

\section{Calculation of Coefficients of Anomalies}

As is well known, the anomalies that can occur
in a symmetry in field theory correspond to the local integrated polynomials
with ghost charge one that lie in the BRS cohomology
space of the BRS operator corresponding to the symmetry.
These polynomials cannot be removed by adding renormalization counterterms.
So once one knows the BRS cohomology of a theory, one knows the possible
anomalies that can arise in that theory.

As mentioned above, the
BRS cohomology of rigid N=1, D=4 supersymmetry is very
rich and complex--it contains many potential anomalies (`holes')
in the BRS cohomology \ci{prl} \ci{cmp2}.

The only set of holes that have been studied so far are those
which are coupled to a spin $\fr{1}{2}$ antichiral superfield.
A completion of the BRS
analysis would probably reveal that there are actually potential anomalies
special to supersymmetry for all half integer spins $\fr{1}{2},
\fr{3}{2},
 \fr{5}{2} \ldots $, and that these anomalies mix these operators
with operators of all integer spins $0,1,2 \ldots $

In order to truly qualify these holes as
anomalies however, it is necessary that the relevant coefficients be
calculated and that at least some of them are not zero. In this section we
illustrate the general situation by
trying to calculate a simple example.

One thing which is clear from the outset is that no supersymmetry
anomalies
can occur unless there are chiral terms in the action of the general
form:
\[
S =  \int d^4 x\; d^2 { \theta} \;
\Bl [  \fr{m}{4} e_{ij} S_i S_j + \fr{1}{6} g_{ijk} S_i S_j S_k
\Br ]
\]
\be
+
 \int d^4 x\; d^2 {\ov \theta} \;
\Bl [ \fr{m}{4}  {\ov e}_{ij} {\ov S}_i {\ov S}_j
+ \fr{1}{6} {\ov g}_{ijk} {\ov S}_i {\ov S}_j {\ov S}_k
\Br ]
\la{chiral}
\ee
where either $e$ or $g$ is not zero.  Here $m$ is a parameter
with the dimension of mass, and ${\ov e}$ is the complex
conjugate of $e$ etc. $S$ are chiral superfields ($ {\ov D}_{\dot \a}
S =0 $) labelled by an
index $i$, which might be an isopin Yang-Mills index.
These terms are needed because
in all examples of possible supersymmetry anomalies
it is necessary to start with chiral fields
and end with antichiral fields and the only way to convert one
to the other is with these chiral terms in the action.

So let us start with the above and the kinetic terms:
\be
S =   \int d^4 x\; d^4 { \theta} \;
\fr{1}{4} [ {\ov S}_i S_i]
\la{kinetic}
\ee

The simplest example of a potential anomaly in this theory arises
when we add the following term to the action:
\be
\int d^4 x\; d^4 { \theta} \;
\fr{1}{8 m} e^{ijk} \F^{\a}_i  S_j D_{\a}  S_k
=
\int d^4 x\; d^2 {\ov \theta} \;
\fr{1}{8 m} e^{ijk} \F^{\a}_i  D^2 (S_j D_{\a}  S_k)
\la{action1}
\ee
Here $\F_{\a}$ is the massive antichiral spinor superfield
discussed at length in \ci{prl}.
For present purposes it can be regarded as an
external superfield source subject to the (antichirality) constraint:
$ D_{\a} \F_{\b i} = 0$.
Here and below we assume that $\F_{\a}$ has its canonical dimension
of $\fr{1}{2}$ and we add the appropriate power of $m$ to make the
other coefficients like $e^{ij}$ dimensionless.  The fact that
a negative power of $m$ is needed in (\ref{action1}) shows that this
coupling is non-renormalizable.  The canonical dimension ($\fr{1}{2}$) of
$\F_{\a i}$ is determined from its kinetic action, which is
discussed in \ci{prl}.

According to the BRS cohomology,
this coupling could give rise to anomalies  of the form:
\be
\left [ c^{\a} Q_{\a} + {\ov c}^{\dot \a}
{\ov Q}_{\dot \a} \right ]
\G \equiv
\d \G = \sum H_i
\ee
where ($c_{\a}$ is the constant supersymmetry `ghost')
\be
H_2 =  \int d^4 x\; d^2 {\ov \theta} \;
a_2^{ij} m^2 \F^{\a}_i c_{\a}  {\ov S_j}
\ee
\be
H_1 = \int d^4 x\; d^2 {\ov \theta} \;
a_1^{ijk} m \F^{\a}_i c_{\a}  {\ov S_j} {\ov S_k} + \cdots
\ee
\be
H_0 =
 \int d^4 x\; d^2 {\ov \theta} \;
a_0^{ijkl}  \F^{\a}_i c_{\a}  {\ov S_j} {\ov S_k} {\ov S_l}
\ee
\be
H_{-1} = \int d^4 x\; d^2 {\ov \theta} \;
a_{-1}^{hijkl} \fr{1}{m}
 \F^{\a}_h c_{\a}   {\ov S_i} {\ov S_j}{\ov S_k} {\ov S_l}
\la{hole-1}
\ee

The `hole' $H_2$, for example, can come from the variation of
the following non-local $\G$:
\[
\G_{2 - \rm anom} =
\fr{1}{4} a_2^{i j} m^2 \int d^4 x \;  d^4 { \theta}  \;
\F^{\a}_i \fr{{\ov D}^2 }{\Box^2} \s^{\m}_{\a {\dot \b} } \pa_{\m}
{\ov Q}^{\dot \b}
 {\ov S}_j
\]
\[
=
- \fr{1}{4}  a_2^{i j} m^2 \int d^4 x \;  d^4 { \theta}  \;
\F^{\a}_i \fr{ \q^2 }{\Box} \s^{\m}_{\a {\dot \b} } \pa_{\m}
{\ov Q}^{\dot \b}
 {\ov S}_j
\]
\be
=  a_2^{ij} m^2 \int d^4 x \;   d^2 {\ov \q}  \; \Bl \{
\F^{\a}_i \fr{1 }{\Box} \s^{\m}_{\a {\dot \b} } \pa_{\m} {\ov Q}^{\dot \b}
{\ov S}_j
\Br \} {|}
\ee
Clearly this term violates supersymmetry and introduces
a non-supersymmetric mass spitting between the two
fields.  If the theory gives rise to such a term, supersymmetry is
violated.

In terms of components, this has the form:
\be
\G_{2- \rm anom} =    2 a_2^{ij} m^2
\int d^4 x\;
[ W^{\m}_i \fr{\pa_{\m}}{\Box}  {\ov F}_j
+
\c^{\a}_i \s^{\m}_{\a \dot \b} \fr{\pa_{\m}}{\Box} {\ov \y}^{\dot \b}_j ]
\ee
where we define  $ W^{\m} = W^{\a \dot \b} \s^{\m}_{ \a \dot \b}$  and
\be
\F_{\a}(x) = \f_{\a}({\ov y})
+ W_{\a \dot \b}({\ov y})  {\bar \q}^{\dot \b}
+ \fr{1}{2} \c_{\a}({\ov y}) {\bar \q}^{\dot \b}
{\bar \q}_{\dot \b}
\ee
\be
S(x) = A(y) + \q \cdot \y(y)
+ \fr{1}{2} \q^2 F(y)
\ee
and  the variable
$ {\ov y}^{\m} = x^{\m} - \fr{1}{2}
 \q^{\a} \s^{\m}_{\a \dot \b} {\bar \q}^{\dot \b}$
is the solution to
\be
D_{\a} {\ov y}^{\m} =
{\ov D}_{\dot \a} y^{\m} = 0.
\ee
Note that an interesting and novel feature of the present situation is
that it is possible that anomalies might arise that have
this mass parameter in them as indicated above.  This does not happen with
the known chiral anomalies, which have parameters like $e_i$ only.
It does happen in ten-dimensional gravity in a sense however.  The
difference is that in gravity (and also non-Abelian Yang-Mills)
there are only a few holes corresponding to the chiral anomalies  for
a given dimension of spacetime, whereas here we have an infinite
number of possibilities.

The relevant part of the action in components is:
\[
S = \int d^4 x \left \{  A_i \Box {\ov A}_i + \y^{\a}_i \s^{\m}_{\a
\dot \b} \pa_{\m} \y^{\dot \b}_i  + F_i {\ov F}_i
\right.
\]
\[
+ \fr{m}{2} \left ( 2 A_i F_i
+   \y_i \cdot \y_i \right )
+ g^{ijk} [  A_i A_j F_k
+ A_i \y^{\a}_j
\y_{\a k} ]
\]
\[
 + \fr{1}{m} e^{i[jk]}
\Bl [ \c^{\a}_i F_j \y_{k \a }
-  W^{\a \dot \a }_i (  {\bar \s}^{\m}_{\dot \a \b}
\pa_{\m} \y^{\b}_j \y_{ k \a }
-  { \s}^{\m}_{\a \dot \a}
F_j \pa_{\m} A_{ k}
\Br )
\]
\be
+
\f^{\a}_i \pa^{\m}
( \pa_{\m} A_j \y_{k \a } )
\Br  ] + {\rm c.c.}
 \Br \}
\ee

The next stage is to see what the BRS identity predicts
for the various two point functions that could give rise
to the anomaly.  A convenient form of the BRS identity is:
\[
\int d^4 x \left \{ c^{\a} \y_{\a i} \fr{\d \G}{\d A_i}
+  [-  F_i c^{\a} + \pa_{\m} A_i
\s^{\m}_{\a \dot \b} {\ov c}^{\dot \b}]
 \fr{\d \G}{\d \y_{\a i} }
\right.
\]
\[
-    \pa_{\m} \y^{\a}_{ i}
\s^{\m}_{\a \dot \b} {\ov c}^{\dot \b}
 \fr{\d \G}{\d F_i }
\]
\[
+ {\ov c}^{\dot \b} W_{\a {\dot \b} i}
\fr{\d \G}{\d \f_{\a i} }
+  [ - \c^{\a}_i {\ov c}^{\dot \b} + \pa_{\m} \f_{\a  i}
(\s^{\m})^{ \b \dot \b} {c}_{\b} ]
\fr{\d \G}{\d W_{\a {\dot \b} i} }
\]
\be
\left.
- \pa_{\m} W^{\a {\dot \b} i}
\s^{\m}_{\b \dot \b} {c}^{\b}
 \fr{\d \G}{\d \c^{\a}_i }
 + {\rm c. c.} \right \}
= H_2
\ee
where the component form of $H_2$ is:
\be
H_2 = 2 m^2 a_2^{ij} \int d^4 x \{ \c^{\a}_i {\ov A}_j
- W^{\a \dot \b}_i {\ov \y}^{\dot \b}_j + \f^{\a}_i {\ov F}_j
\} c_{\a}
\ee
and from this we can deduce,
 using functional differentiation and
Fourier transform,
identities
 in momentum
space relating various n-point functions
to the anomalies (if they are present).

So, for example, taking functional derivatives
with respect to $ \ov F, \f_{\a}$ and $c_{\a}$, we get:
\be
k^{\m} \G_{ij  \m}({\ov F} , W)
=
4 a_{ij} m^2
\ee
where a is the coefficient of the anomaly,
\be
\G_{ij  \m}({\ov F} , W)
= \fr{\d^2 \G}{\d {\ov F}_i \d W^{\m}_j } |
\ee

So, for this example,
 we only have to calculate one 1PI diagram:

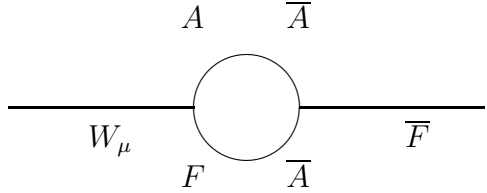
\begin{picture}(400,150)
\put(85,75){\line(1,0){70}}
\put(115,60){$W_{\m}$}
\put(150,105){$A$}
\put(150,45){$F$}
\put(190,45){${\ov A}$}
\put(175,75){\circle{40}}
\put(190,105){${\ov A}$}
\put(235,60){${\ov F}$}
\put(195,75){\line(1,0){70}}
\end{picture}

This diagram is clearly zero because
it would require a propagator which converts $F_i$ into
${\ov A}_j$ and this does not exist in the theory as it stands.
The relevant bosonic kinetic term is:
\be V^{\dag} K V =
\left(  {\ov A}_i, A_i,
{\ov F}_i, F_i ] \right)
\left( \ba{cccc} - k^2 &  0 & 0 & m \\ 0 & - k^2 & m & 0  \\
 0 & m & 1  & 0 \\ m & 0 & 0 & 1 \ea
\right) \left( \ba{c} A_i\\ {\ov A}_i\\
 F_i\\ {\ov F}_i \ea \right)
\ee
whose inverse is:
\be K^{-1} =
\left( \ba{cccc} - \fr{1}{k^2 + m^2}  &  0 & 0 & \fr{m}{k^2 + m^2}
 \\ 0 & - \fr{1}{k^2 + m^2}
& \fr{m}{k^2 + m^2}
& 0  \\
 0 & \fr{m}{k^2 + m^2}
 & \fr{k^2}{k^2 + m^2}
  & 0 \\ \fr{m}{k^2 + m^2}
 & 0 & 0 &  \fr{k^2}{k^2 + m^2}
\ea
\right)
\ee
which clearly has no term that can give rise to the
indicated two point functions.  Diagrams with different
component fields behave in a similar way.

Since there are no diagrams that can contribute to the
relevant two point functions, it would seem unnecessary
to regulate these amplitudes, and hence no anomaly should develop.
This kind of result was behind the arguments made in \ci{prl} that
no anomalies should arise for this cohomology--it seems to be quite
general in fact.
But is this computation really correct?  Note that if one calculates the
usual axial anomaly naively in an unregulated theory, one incorrectly
obtains zero.  So at the very least we should regulate the theory.
But how?  The most reasonable suggestion might be to try `supersymmetric
Pauli-Villars regularization' in superspace--that ought to
show that there are no supersymmetry anomalies, at least when it can be used.
But even supersymmetric Pauli-Villars for chiral
superfields  is in doubt
for three reasons: (1) massless fields cause a problem, (2) when
implemented with a regularized action, Pauli-Villars requires the wrong
connection between spin and statistics for the regulator fields,
 which in turn
plays havoc with
the supersymmetry algebra--the Hamiltonian is no longer positive
definite, which means that $\sum_i Q_i^2 = H$
must be modified somehow. (3) Pauli-Villars breaks gauge invariance
when supersymmetric Yang-Mills is present.

My guess is that the result is correct, in spite of these worries.
  However it is much less
clear what happens at higher orders in a theory with Yang-Mills and
chiral matter and spontaneous gauge breaking.  To keep the Yang-Mills
symmetry manifest, one would like to use dimensional regularization.
But it is clear that dimensional regularization is very hard
to reconcile with supersymmetry, and that the problems start to show
up at multiloop orders.
It is also easy to convince oneself that some anomalies could not possibly
arise at one loop simply because it requires more than one loop to
even generate the appropriate outgoing fields.  For example, one could
take:
\be
\Y_{\a} = D^2 \{ U^i_{L}
D^j_{L}
[e^{-V} D_{\a} e^{V} U_{L}]^k
\e_{ijk} \}
\ee
as the type of operator that could create a proton.  It requires two loops
to convert all the chiral fields to antichiral ones, so could not
possibly have a supersymmetry anomaly before two loops.

Similarly one could make an operator to create an electron from
the Higgs chiral superfield and the electron matter chiral superfield.
The VEV of the Higgs would allow the electron flavour to
`emerge'.  Again a supersymmetry anomaly could split its mass from the mass
of its superpartners.

\section{Conclusion}

The non-trival BRS cohomology of N=1, D=4 supersymmetry
naturally gives rise to a number of speculations concerning
the origin of the dimension of spacetime, the uniqueness of
our own universe and the reason that supersymmetry breaking
occurs with a zero cosmological constant.  It would be
interesting to determine whether the known supersymmetry
cohomology carries over
to local supersymmetry and higher dimensions, and it is
desirable that the coefficients of some of the anomalies be
calculated.  It would also be interesting to work out the
supersymmetry breaking
consequences of the anomalies, assuming that they are indeed present
at some order of perturbation theory.

How can one compute the coefficients of the holes in supersymmetric theories?
In order for the theory to exhibit supersymmetry anomalies,
one must be forced to use a regularization which explicitly violates
supersymmetry.  Indeed, this appears to be happen if one insists on
regularizing in a way that preserves gauge symmetry, which is exactly
the case of interest for phenomenology.

\end{document}